\documentclass[a4paper,11pt]{article}
\usepackage{pos}

\newcommand{\gettitle}{}
\hypersetup{linkcolor=black
	colorlinks,
	linkcolor={red!75!black},
	citecolor={blue!75!black},
	urlcolor={blue!75!black},
	pdftitle={\gettitle},
	pdfauthor={Barata, Mehtar-Tani},
	pdfkeywords={Perturbative QCD} {Jet quenching},
	bookmarksopen=true,
	bookmarksopenlevel=2,
	bookmarksnumbered=true
}
\makeatletter
\newcommand\makebig[2]{%
  \@xp\newcommand\@xp*\csname#1\endcsname{\bBigg@{#2}}%
  \@xp\newcommand\@xp*\csname#1l\endcsname{\@xp\mathopen\csname#1\endcsname}%
  \@xp\newcommand\@xp*\csname#1r\endcsname{\@xp\mathclose\csname#1\endcsname}%
}
\makeatother
\makebig{biggg} {1.3}

\def\bs{\boldsymbol} 
\def\del{\partial}
\def\bdel{\bs\partial}

\newcommand{\eqn}[1]{Eq.~\eqref{#1}}

\long\def\comment#1{ }

\newcommand{\nn}{\nonumber\\ }

\def\be{\begin{eqnarray*}}
\def\ee{\end{eqnarray*}}
\def\beq{\begin{eqnarray}}
\def\eeq{\end{eqnarray}}
\newcommand{\bea}{\beq \begin{aligned}}
\newcommand{\eea}{\end{aligned}\eeq}

\def\k{{\boldsymbol k}}

\def\q{{\boldsymbol q}}

\def\x{{\boldsymbol x}}
\def\y{{\boldsymbol y}}
\def\z{{\boldsymbol z}}

\def\0{{\boldsymbol 0}}
\def\l{{\boldsymbol l}}
\def\k{{\boldsymbol k}}

\def\x{{\boldsymbol x}}
\def\y{{\boldsymbol y}}

\def\z{{\boldsymbol z}}

\def\lambdaq{\lambda}

\def\rme{{\rm e}}
\def\dd{\text{d}}

\def\and{ \quad\text{and}\quad}

\def\Re{\text{Re}}

\def\cP{{\cal P}}
\def\cK{{\cal K}}

\def\rmd{{\rm d}}

\def\dd{\text{d}}

\def\dd{\text{d}}

\title{Medium induced gluon spectrum in the Improved Opacity Expansion}

\author*[a]{João Barata}

\affiliation[a]{Instituto Galego de Fisica de Altas Enerxias (IGFAE), Universidade de Santiago de Compostela,E-15782 Galicia, Spain}

\emailAdd{joaolourenco.henriques@usc.es}

\abstract{Over the last decades, analytical calculations of jet quenching observables have always needed to make a distinction between dense or dilute mediums. Although there are different theoretical formalisms suited for each one of these scenarios, taking into account multiple soft and single hard interactions between the probe and the background under a single approach has proven to be a difficult task. In this talk, we will introduce the Improved Opacity Expansion (IOE), which extends the well established Opacity Expansion framework beyond the hard momentum transfer tail, including the regime captured by the BDMPS-Z/ASW approximation. We will focus on the application of the IOE to the computation of the single gluon medium induced spectrum from a hard parton, which constitutes one of the most important theoretical results in jet quenching theory.}

\FullConference{%
  *** Particles and Nuclei International Conference - PANIC2021 ***\\
  *** 5 - 10 September, 2021 ***\\
  *** Online ***
}

\begin{document}
\maketitle

\section{Introduction}
Jet quenching is one of the main signatures for the creation of a quark gluon plasma after the collision of heavy ions at RHIC and the LHC~\cite{ Adcox:2001jp, CMS:2012aa}. Broadly, it corresponds to the modification of the jet's structure due to the interactions with the underlying hot medium, resulting in the jet's constituents broadening their momentum due to scattering with the medium quasi-particles and the production of induced bremsstrahlung radiation. In the past decades, analytical calculations of these effects have focused on the dilute and dense medium limits, where exact analytic results can be derived~\cite{BDMPS1,BDMPS2,BDMPS3,BDMPS4,BDIM1,Arnold:2020uzm,Gyulassy:2000fs,Wiedemann,Guo:2000nz,Vitev:2009rd,Chien:2015hda,GLV}. Even though dedicated numerical routines have been designed to study such processes~\cite{numerical1,Zakharov:2004vm,numerical2,numerical3,CarlotaFabioLiliana,Andres:2020kfg,Feal:2019xfl}, for many phenomenological applications it would be valuable to have semi-analytical methods to connect these two limits.

In this proceedings, we discuss a novel strategy, dubbed the Improved Opacity Expansion (IOE), which allows to merge the thin and dense medium limits under a single framework. The general strategy of the IOE is first illustrated for the evolution of a single particle in the medium (single particle momentum broadening) and is then used to compute the medium induced gluon spectrum.

\section{A working example: momentum broadening in the IOE approach}
To introduce the IOE formalism~\cite{IOE_b,IOE1,IOE2,IOE3,IOE4} we consider first the simplest possible observable: single particle momentum broadening. At eikonal accuracy, momentum broadening can be fully characterized by an associated broadening distribution $\cP(\k,t)$ which gives the probability for a parton to acquire transverse momentum $\k$ in the medium as a result of its propagation for a time $t$. In terms of the in-medium elastic collision rate $\gamma_{\rm el}(\q)\sim  g^4n/\q^{4}$, it reads~\cite{BDIM1} 
\beq
\label{eq:rate-eq}
 \frac{\del  \cP(\k,t)}{\del  t } = C_R\int_\q \,\gamma_{\rm el}(\q) \left[\cP(\k-\q,t) -\cP(\k,t) \right]\, ,
\eeq
where $C_R$ is the associated Casimir color factor and $n$ is the density of in-medium scattering centers.

For thin mediums, i.e. when the hard parton interactions a small number of times with the background, $\cP$ is dominated by at most a single hard scattering (SH) and it reduces to 
\beq\label{eq:oe-lo-br}
 \cP^{\rm SH}(\k,L)= C_R\gamma_{\rm el}(\k)\,  L \propto   \frac{\alpha_s^2C_R nL}{\k^4}\, .
\eeq
Conversely, when the medium is dense the probability of scattering becomes of order one and all possible (soft) tree-level gluon exchanges with the medium have to be taken into account. In this regime of multiple soft (MS) scattering with the medium, \eqn{eq:rate-eq} can be reduced to a diffusion equation, so that the leading order result reads 
\beq\label{eq:gaussian}
  \cP^{\rm MS}(\k,L) =\frac{4\pi }{\hat q  L } \rme^{-\frac{\k^2}{
  \hat q L }} \, ,
\eeq
where $\hat q$ is the jet quenching parameter  
\beq 
\hat q = C_R\, \int^{q_{\rm max}}_{\mu_\ast}\frac{\rmd^2 \q }{(2\pi)^2} \, \q^2 \, \gamma_{\rm el}(\q) \approx 4\pi \alpha_s^2 C_R n \log\frac{q_{\rm max}^2}{\mu_\ast^2} \equiv \hat q_0 \log\frac{q_{\rm max}^2}{\mu_\ast^2}\, .
\eeq
Notice that the in the infrared this integral converges due to the thermal mass ($\mu_\ast$) screening and in the ultraviolet the logarithmic divergence must be regulated by a large momentum scale $q_{\rm max}$. Comparing \eqn{eq:oe-lo-br} and \eqn{eq:gaussian}, it is clear that the SH solution properly describes the expected $1/\k^4$ Coulomb form at large momentum, while the MS solution recovers the typical Gaussian solution associated to the resummation of multiple exchanges of soft instantaneous gluons. Although both formulas cover the full range of phenomenological interest for jet quenching when combined, each approximation fails to properly describe the underlying physics in the regime of validity of the other.

\begin{figure}[t!]
  \centering
  \includegraphics[scale=.7]{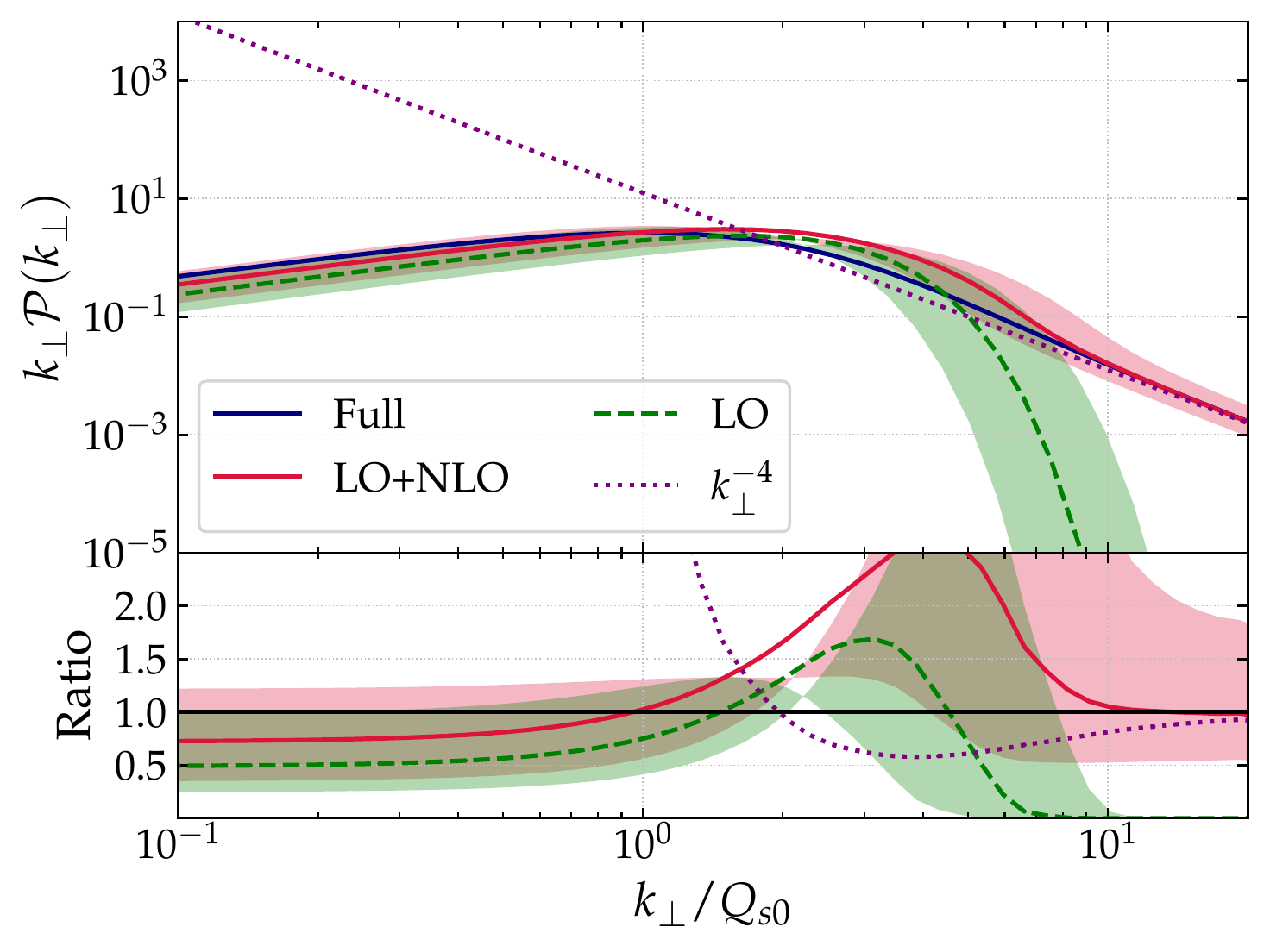}
  \caption{Comparison between the broadening probability distribution in the IOE at LO (dashed, green), at LO+NLO (solid, red) and the exact Gyulassy-Wang model~\cite{GW} result (solid, navy).The medium parameters used in this proceedings are $\hat q_0\!=\!0.16$~GeV$^3$, $L=6$~fm and $\mu_\ast=0.355$~GeV. See~\cite{IOE4} for more details.}
  \label{fig:broad}
\end{figure}

The goal of the IOE expansion is to bridge the gap between the SH and MS regimes. For that we note that Fourier transforming \eqn{eq:rate-eq} gives
\beq\label{eq:rate-position}
 \frac{\del  \cP(\x,t)}{\del  t } = - \, v(\x) \cP(\x,t) \,,
\eeq
with 
\beq \label{eq:v-llog}
v(\x) = C_R\int_\q \, \gamma_{\rm el}(\q)  \left(1- \rme^{i\q\cdot \x}\right) \propto \x^2 \log \frac{1}{\x^2 \mu_\ast^2 }+ \mathcal{O}(\x^4\mu_\ast^2) \, ,
\eeq
where in the last term we used the fact that at sufficiently small distances the interaction potential is assumed to be of the Coloumb form, and thus any medium model should recover \eqn{eq:v-llog} in that regime. Also, by direct comparison with the above MS result, we observe that \eqn{eq:gaussian} is recovered by freezing the logarithm in $v(\x)$ (via an extra scale $q^2_{\rm max}$).
To combine both these regimes, in the IOE, one introduces a matching scale $Q_b^2$ to rewrite $v(\x)$, such that
\begin{equation}\label{eq:ppp}
  v(\x)\propto \x^2 \log \frac{1}{\x^2 \mu_\ast^2 } = \x^2 \left[ \log \frac{Q_b^2}{ \mu_\ast^2 }  + \log \frac{1}{\x^2 Q_b^2 }  \right] \propto v_0 (\x)+\delta v(\x) \, .
\end{equation}
The first term recovers the MS solution (assuming $Q_b^2=q^2_{\rm max}$). Requiring that $Q_b^2\gg\mu_\ast^2$, the second term only becomes dominant at small $\x$ (large $\k$), and thus it can be treated perturbatively. As a result the broadening distribution can be written as~\cite{IOE_b}
\begin{equation}\label{eq:expli_cP_series}
  \begin{split}
  \cP(\k,L)&=\int_\x \, \rme^{-i \x \cdot \k }\rme^{-\frac{1}{4}\x^2Q^2}\sum_{n=0}^{n_{\rm max}}  \, \frac{(-1)^n Q_{s0}^{2n}}{4^nn!} \, \x^{2n} \log^{n}\frac{1}{\x^2Q_b^2}\equiv \cP^{\rm LO}(\k,L) + \cP^{\rm NLO}(\k,L) + \ldots \, ,
  \end{split}
  \end{equation}
where $Q^2_{s0}=\hat q_0 L$ and the matching scale must obey the recursive relation 
\begin{align}\label{eq:old_Qb}
  &Q^2_b\equiv Q^2_{s0} \,\log \frac{Q_b^2( L)}{\mu_\ast^2}\, ,
 \end{align}
in order to eliminate spurious divergences (see~\cite{IOE3,IOE_b} for a detailed discussion). The first two terms in the previous power series can be computed exactly:
\begin{equation}\label{eq:golden}
  \cP^{\rm{LO+NLO}}(\k,L)=  \frac{4\pi}{Q_s^2} \rme^{-x} - \frac{4\pi}{Q_s^2} \lambdaq \left\{1-2 \rme^{-x} + \left(1-x\right) \left[{\rm Ei}\left( 4x\right)-\log 4x\right] \right\}\, ,
  \end{equation}
  where $x = \k^2/Q_s^2=\k^2/(\hat q_0 L \log \frac{Q_b^2}{\mu_\ast^2})$ and $\lambda = \log^{-1 }\frac{Q_b^2}{\mu_\ast^2}\ll 1$. From the above discussion, one expects that perturbative term $\cP^{\rm NLO}$ correctly describes the hard $\k$ tail, while the LO contribution is easily seen to recover the MS result. Indeed, a short calculation shows that at large momentum the NLO becomes dominant and reduces to \eqn{eq:oe-lo-br}:
  \begin{align}\label{eq:plot_ppp}
    \cP(\k,L)^{\rm{NLO}}\Big\vert_{\k^2\gg Q_{s}^2}=4\pi\frac{Q_{s0}^2}{\k^4}+\mathcal{O}\left(\frac{Q_{s0}^4}{\k^6}\right) \, .
    \end{align}
  In Fig.~\ref{fig:broad}, we compare \eqn{eq:golden} to an exact numerical evaluation of $\cP$ for a realistic $v(\x)$ (see~\cite{IOE_b,IOE4} for details), explicitly confirming the above observations.

\section{Medium induced spectrum in the IOE approach}
In this section we discuss how to apply the IOE approach, illustrated for the broadening case above, to the medium induced (soft) gluon spectrum, differential in the gluon's energy $\omega$ and transverse momentum $\k$. In general, this can be written as~\cite{IOE4}
\begin{align}\label{eq:spectrum-2}
  (2\pi)^2\omega\frac{\rmd I}{\rmd \omega  \rmd^2 \k}&=\lim_{\epsilon\to 0}\frac{2\bar{\alpha}\pi}{\omega^2} \Re \int_0^\infty \rmd t_2 \, \rme^{- \epsilon t_2} \int_0^{t_2}  \rmd t_1 \int_\x \,  \rme^{-i \k \cdot \x}\, \cP(\x,\infty;t_2) \nn
  & \bdel_\x\cdot \bdel_\y \cK(\x,t_2;\y,t_1)_{\y=0}  - \frac{8\bar{\alpha}\pi}{k_\perp^2} \,.
\end{align}
Notice that the previous formula only takes into account purely medium induced radiation off the nascent parton and the adiabatic phase $\rme^{- \epsilon t_2} $ ensures that interactions are properly turned-off at asymptotically large times. In the present form, one observes that the full spectrum can be written as the broadening of a gluon final state in the time interval $(t_2,\infty)$, being preceded by its production via an effective kernel $\cK$ during $(t_1,t_2)$. 

The treatment of $\cP$ follows the discussion in the previous section. To deal with the kernel $\cK$, we use the fact that it obeys the following Dyson equation~\cite{Arnold_simple,IOE1}
\begin{align}
  \label{eq:k-expansion}
  \cK(\x,t_2;\y,t_1)  = \cK^{\rm LO}(\x,t_2;\y,t_1)-\int_\z \int_{t_1}^{t_2} \dd s \, \cK^{\rm LO}(\x,t_2;\z,s)\delta v(\z,s)\cK(\z,s;\y,t_1) \, ,
  \end{align}
where $\delta v$ is given in \eqn{eq:ppp}, up to an overall normalizing factor. Truncating \eqn{eq:k-expansion} to first order in $\delta v$, we can write \eqn{eq:spectrum-2} to first order in the IOE as 
\beq
\label{eq:spectrum-ioe}
\frac{\rmd I}{ \rmd \omega  \rmd^2 \k} = \frac{\rmd I^{\rm LO}}{ \rmd \omega  \rmd^2 \k} + \frac{\rmd I^{\rm NLO}}{ \rmd \omega  \rmd^2 \k} + \mathcal{O}(\delta v^2) \,.
\eeq
The first term corresponds to the MS result with a matching scale $Q_r$ given by (see \cite{IOE4,IOE3} for details)
\begin{equation}\label{eq:Q_r}
  Q_r^2=\sqrt{\hat{q}_0\omega\log{\frac{Q_r^2}{\mu_\ast^2}}} \, ,
  \end{equation}
where $\sqrt{\hat q_0 \omega}$ is the typical transverse momentum accumulated by the gluon during the emission process. The NLO term takes into account the leading order corrections in $\delta v$. Explicitly these it can be written as
\begin{align}
  \label{eq:spectrum-ioe-nlo}
  (2\pi)^2\omega\frac{\rmd I^{\rm NLO}}{ \rmd \omega  \rmd^2 \k}& = \frac{2\bar{\alpha}\pi}{\omega^2} \Re \int_0^\infty \rmd t_2\, \rme^{-\epsilon t_2} \int_0^{t_2} \rmd t_1 \int_\x\, \rme^{-i \k \cdot \x}  \Big[ \cP^{\rm LO}(\x,\infty;t_2) \bdel_\x\cdot \bdel_\y \cK^{\rm {NLO}}(\x,t_2;\y,t_1)_{\y=0}  \nn
  &+ \cP^{\rm NLO}(\x,\infty;t_2) \bdel_\x\cdot \bdel_\y \cK^{\rm {LO}}(\x,t_2;\y,t_1)_{\y=0} \Big]\,,
  \end{align}
where we used  $ \cP^{\rm NLO}(\x, \infty;t) = - \cP^{\rm LO}(\x,\infty;t) \int_t^\infty \rmd s \, \delta v(\x,s)$ and the first order truncation of \eqn{eq:k-expansion}: $\cK^{\rm NLO}(\x,t_2;\y,t_1) = - \int_\z \int_{t_1}^{t_2} \rmd s\, \cK^{\rm  LO}(\x,t_2;\z,s) \delta v(\z,s) \cK^{\rm LO}(\z,s;\y,t_1)\, . $ Notice that the first order correction to the MS result given in \eqn{eq:spectrum-ioe-nlo} has a term related to modification of the emission kernel (followed by the usual Gaussian broadening), and a term describing the production of a gluon in the MS approximation followed by broadening dominated by single hard scattering contribution. Analogously to the exercise done in the previous section, one can show that the IOE exactly recovers the SH scattering behavior at large gluon energy and transverse momentum. Performing a slightly evolved calculation, it results that (see details in~\cite{IOE4})
\begin{align}\label{eq:IOE_large_kt}
  &(2\pi)^2\omega\frac{\rmd I^{\rm NLO}}{\rmd \omega \rmd^2 \k}\approx\frac{8\pi \bar{\alpha}\hat q_0L}{\k^4}\left[3 \gamma_E-4+\log\left(\frac{\k^2}{4Q_b^2}\right)+\log\left(\frac{\k^2L}{2\omega}\right)\right]\,.
  \end{align}
Unlike in \eqn{eq:plot_ppp}, it seems that there is still a dependence on the matching scale $Q_b$. However, this dependence vanishes once LO subleading corrections are taken into account. This feature, i.e. the independence of the IOE on the matching scales at high energies, seems to be generic. The full spectrum at this accuracy exactly gives
\begin{align}
  \label{eq:final-highenergy}
    (2\pi)^2\omega\frac{\rmd I^{\rm LO+NLO}}{\rmd \omega \rmd^2 \k}&= \frac{8\pi\bar{\alpha}\hat{q}_0L}{ \k^4}\left[3\gamma_E-4+\log\left(\frac{\k^2}{4\mu_\ast^2}\right)+\log\left(\frac{\k^2L}{2\omega}\right)\right]= (2\pi)^2\omega\frac{\rmd I^{\rm SH}} {\rmd \omega  \rmd^2 \k}\,,
\end{align}
as expected.

\begin{figure}[h!]
  \centering
  \includegraphics[scale=.45]{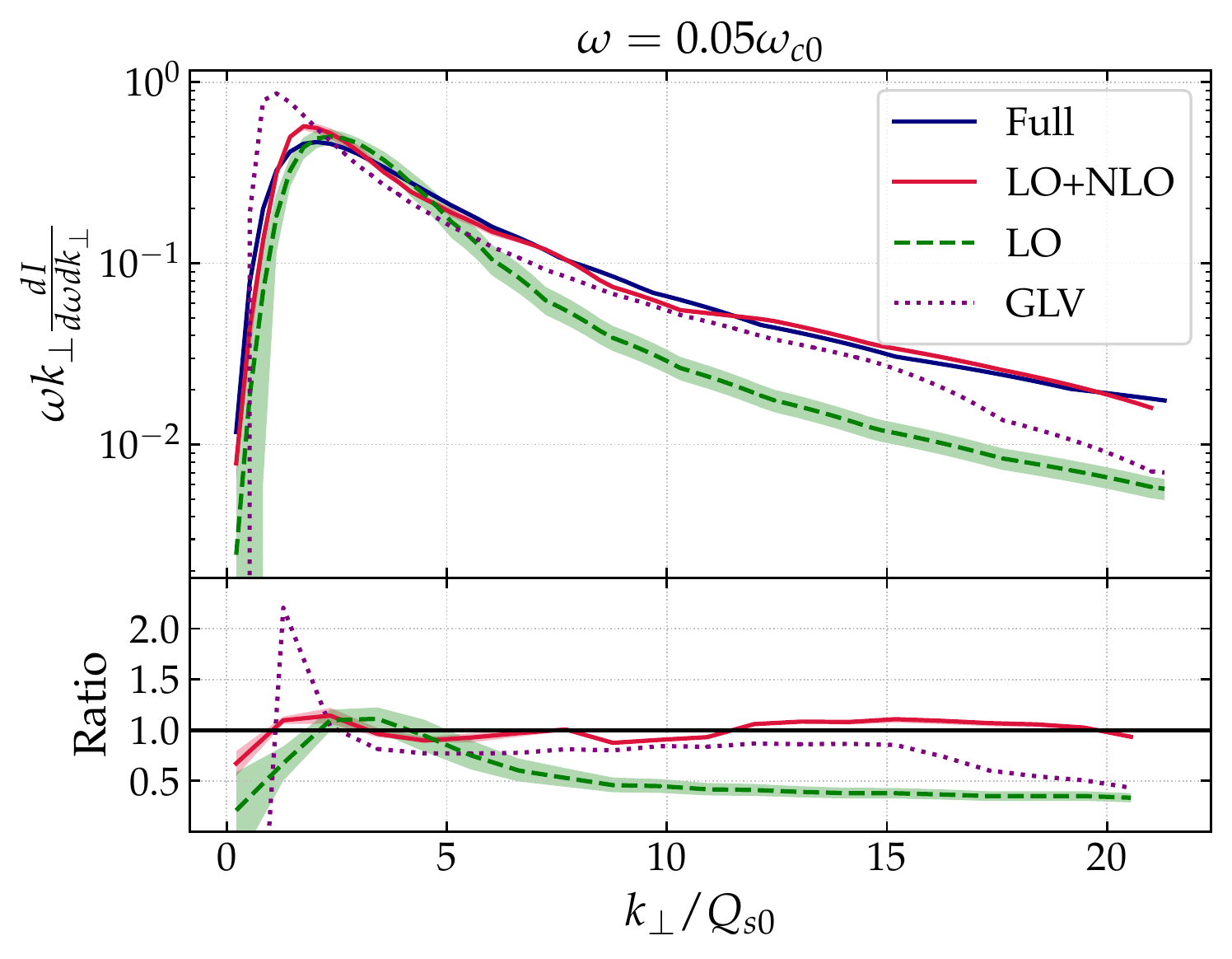}
  \includegraphics[scale=.45]{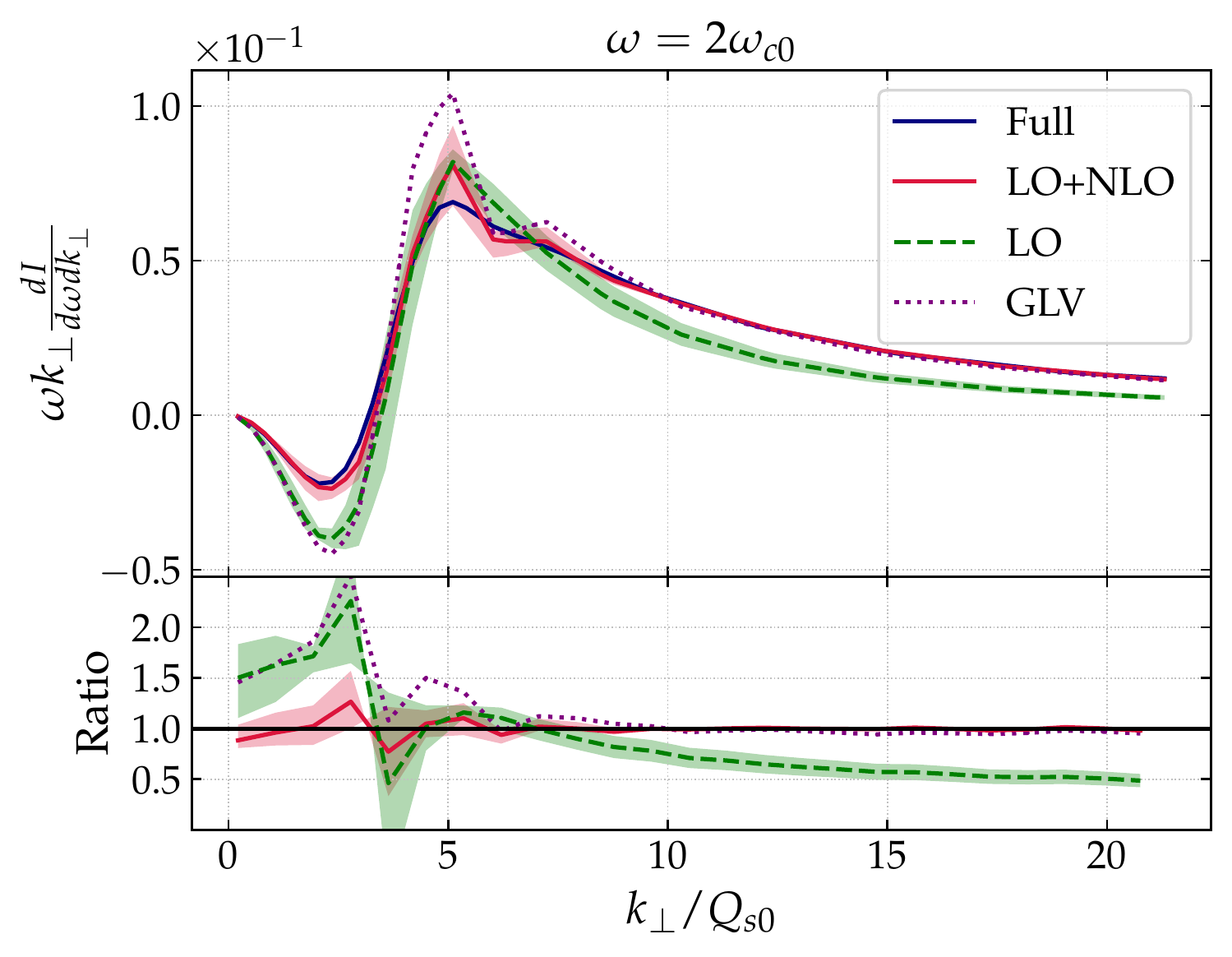}
  \caption{Comparison between the SH (GLV) spectrum (dotted, purple), the LO result (dashed, green), the IOE at LO+NLO (solid, red) and the all-order spectrum (solid, navy) as computed in~\cite{CarlotaFabioLiliana} for two gluon frequencies: $\omega= 0.05\, \omega_{c0}$ (left) and $\omega= 2\, \omega_{c0}$ (right). The ratio to the full solution is presented in the bottom panels. The uncertainty band arises from variations in the matching scales $Q_b$ and $Q_r$.}
  \label{fig:lhc}
\end{figure}

The expressions given above can be further simplified; a numerical code to compute them can be found in \cite{python_git} with comparable computational cost with respect to previous implementations of the MS result~\cite{ASW2}. To illustrate the accuracy of the IOE approach, in Fig.~\ref{fig:lhc} we show the differential spectrum for two gluon frequencies ($\omega_{c0}\equiv \hat q_0 L^2$), in a set up close to LHC conditions. We compare the result to an exact novel numerical solution (see \cite{CarlotaFabioLiliana} for details) and to the SH~\cite{GLV} result, corresponding to the case where there is a single hard scattering center in the medium. At a qualitative level, the IOE leading order expansion already seems to perform rather well, especially when compared to the either the SH or MS (i.e. the LO term in the IOE) approximations.

\section{Conclusion and Outlook}
We have illustrated how to extend the Improved Opacity Expansion to the case of the double differential medium induced gluon spectrum. Along with single particle momentum broadening, these two observables constitute the back-bone  of many phenomenological jet quenching models of full jets. Thus, the IOE formalism allows for the first time to include the effects of both multiple soft and single hard scattering in the medium under a single framework. 

An upside of the IOE approach is that it admits a semi-analytical treatment~\cite{Mehtar-Tani:2021fud, Takacs:2021bpv}. As a result, one can use this formalism in, for example, jet substructure studies, gaining important analytical control over how the MS and SH regimes affect desired observables. In addition, the simple final expressions obtained can be generalized to arbitrary medium time profiles and thus could offer an efficient avenue to implementing jet evolution in dynamical backgrounds.

\section*{Acknowledgements}
The author would like to thank Yacine Mehtar-Tani, Alba Soto-Ontoso and Konrad Tywoniuk for their collaboration. We are also grateful to the authors of~\cite{CarlotaFabioLiliana} for providing the numerical results from their study.
 J.B. is supported by Ministerio de Ciencia e Innovacion of Spain under project FPA2017-83814-P; Unidad de Excelencia Maria de Maetzu under project MDM-2016-0692; European research Council project ERC-2018-ADG-835105 YoctoLHC; and Xunta de Galicia (Conselleria de Educacion) and FEDER.



\bibliographystyle{elsarticle-num}

\bibliography{Lib.bib}

\end{document}